# MODELING WITH STRUCTURE OF RESINS IN ELECTRONIC COMPONENTS

Qiang YU*, Tadahiro SHIBUTANI*, Masaki SHIRATORI*, Tomio MATSUZAKI*

Tsubasa MATSUMOTO*

*Department of Mechanical Engineering and Materials Science

Yokohama National University

79-5, Tokiwadai, Hodogaya-ku, 240-8501, Japan

**ABSTRACT**

In recent years, interfacial fracture becomes one of the most important problems in the assessment of reliability of electronics packaging. Especially, underfill resin is used with solder joints in flip chip packaging for preventing the thermal fatigue fracture in solder joints. In general, the interfacial strength has been evaluated on the basis of interfacial fracture mechanics concept. However, as the size of devices decrease, it is difficult to evaluate the interfacial strength quantitatively. Most of researches in the interfacial fracture were conducted on the basis of the assumption of the perfectly bonding condition though the interface has the micro-scale structure and the bonding is often imperfect. In this study, the mechanical model of the interfacial structure of resin in electronic components was proposed. Bimaterial model with the imperfect bonding condition was examined by using a finite element analysis (FEA). Stress field in the vicinity of interface depends on the interfacial structure with the imperfect bonding. In the front of interfacial crack tip, the behavior of process zone is affected by interfacial structure. However, the instability of fracture for macroscopic crack which means the fracture toughness is governed by the stress intensity factor based on the fracture mechanics concept.

## 1. INTRODUCTION

In recent years everyday life has become more and more convenience through significant progress in computerization. On the other hand, there is great stress put upon the technology that supports these new technologies. There is no end to demanding smaller, lighter, more high tech and faster electronic devices, and the modern field of science is struggling with many problems that need to be overcome. These problems include expanding the variation of forms and materials used in the production, adaptation to a wider range of environment where these technologies may be used, making device forms more complicated, consideration of environmental issues (i.e. reduction of lead usage, recycling, long life expectancy designing), and foundation of a method for reliability designing. One of the significant methods to improve the strength of chip assemblies these days is the dispensing of resin between the spaces between the silicon chip and the substrate, where the circuits come in action through solder balls[1,2]. The most critical failures in chip assembles are the disconnections of the circuit, and the joints of different materials, such as the chips and the solder ball joints, are the most susceptible to failure. These failures occur when stress concentrates at the weak points in the structure often caused by shocks and differences in the coefficient of thermal expansion (CTE) in different materials. Resin materials are used for the purpose of redistributing the stress and reducing the effect of differences in CTE.

Since electronic devices consist of various materials, there are many interfaces: metal/ceramic, metal/resin, ceramic/resin and so on. Therefore, interfacial cracking affects the reliability of devices. Interfacial strength has been often evaluated as the fracture toughness on the basis of fracture mechanics concept[3,4]. Perfect bonding was assumed and stress intensity factor is dominant parameter of interfacial failure. However, as the size of devices decreases, interfacial fracture behavior becomes more complicated. Figure 1 shows an example of interfacial cracking between resins. These materials are bonded partly and interfacial structure can be seen at the crack tip. As the devices size becomes smaller, the effect of interfacial structure appears. Additionally, in the interfacial facture mechanics, the failure process of interface has been ambiguous yet. Therefore, interfacial structure-based evaluation should be constructed. In this study, the simple partly bonded interface model was proposed. FEA for the proposed model was compared with the bimaterial interface model. The effect of interfacial structure was extracted. Considering the failure criteria of the interfacial structure, the behavior of interfacial fracture was examined.

## 2. INTERFACIAL STRUCTURE BASED MODEL

Figure 2 shows the bimaterial model with interfacial structure. Bonding parts of which width and height are *a* and *h*, are arrayed periodically with the interval *s*. FEA analysis was conducted under two boundary conditions: uniaxial tensile and thermal loading. In the former cases as shown in Fig.2, the bottom and right sides were fixed. Uniform tensile stress was applied on the upper of resin. In the latter case, temperature changed from 150 C to 20 C. No stress is applied at 150 C. Five models were prepared where sizes of bonding structure, *a* and *h* range as shown in Table I. 4-node plain strain element was used and their numbers of elements and nodes are shown in Table I. Fine mesh was constructed near the interfacial zone. The solver is Marc 2005. Materials are assumed to be isotropic elastic body and each material property is listed in Table II. Additionally, by considering the fracture property of interfacial structure, the failure behavior at the crack tip was investigated. In this case, the interfacial structure and resin is assumed to be elasto-plastic body. In order to determine the criteria of fracture, the maximum plastic strain was employed. Using the subroutine "uactive", elements which reach the criteria are deactivated.

## 3. RESULT AND DISCUSSION

### 3.1. Uni-axial tensile

Figure 3 (a) and (b) are distributions of $\sigma_y$ and $\tau_{xy}$ along the interface, respectively. Each plotted data is taken at the middle of interfacial structure. The stress curves in bimaterial model with perfect bonding are also shown by solid lines. When *s* is small, the curve is close to the solid line. Since separated area is small, the structure is almost that in the perfect bonding. Other curves are higher than the solid line. When interfacial height h is low, the stress $\sigma_y$ increases. The mismatch of deformation concentrates near the interface and the strain in the normal direction depends on the height of interfacial structure. On the other hands, in the case of $\tau_{xy}$, when interfacial height *h* is low, the stress decreases. It suggests that the interfacial structure affects the mix mode condition at the edge of interface. The effect of interface's height appears in the middle of model. Though the horizontal mismatch of displacement is generated at the edge of interface, the vertical mismatch affects the shear stress in the middle of model.

### 3.2. Residual stress

Figure 4 (a) and (b) shows the distribution of the residual stresses. When *a*=5 and *s*=5, (the fraction of bonded area is 50%), the residual stress, $\sigma_y$ increases. Not only the bonded area decreases, and but also the mismatch of deformation is changed due to the rigidity of interfacial structure. It implies that the decrease in the bonding area causes that the effect of interfacial structure appears. When the height of interfacial structure, *h* becomes high, the stress increases. The effective zone of localized deformation due to the mismatch depends on the bonding structure. When *h* is low, not only the interfacial structure but also bulk materials can be deformed by the mismatch of CTE. On the other hand, when *h* is high, the deformation by mismatch concentrates on the bonding area.

Comparing with the uniaxial tensile state, the effect of interface height on the shear stress is not so big. CTE mismatch does not depend on the interfacial structure and subject each interfacial structure. Therefore, the width and interval of bonding part mainly affect the behavior of shear stress.

### 3.3. Fracture of interface

Figure 5 and 6 show the failure process, near the crack tip under uniform tensile stress state. The fracture in the interfacial structure occurred in the front of the tip. As the applied stress increase, failure process proceeds. Finally, instability of fracture takes place. Figure.7 and Figure.8 are the stress distribution from the tip at the instability of fracture. When the stress singularity field near the crack tip reaches critical value, the instability begins. It implies that macroscopic stress field governs the interfacial fracture. On the other side, stress depends on the interfacial structure in $r < 100$ μm. Therefore, the process zone is affected by the interfacial model. In the case of electronics components, the size of failure is less than that of process zone. It suggests the importance of the interfacial structure model concept proposed in this paper.

## 4. CONCLUSION

1) Interfacial stress behavior in the state of elasticity of the proposed two dimension model was able to be confirmed.
2) The stress field along the interface depends on the interfacial structure. At the edge of interface, mixed mode state is affected by the bonding structure.
3) In the front of the crack, the process zone is affected by the interfacial structure. On the other hand, when the crack length is long, the fracture toughness obeys on the stress singularity field along the interface.
4) In electronics components, the fracture behavior of process zone affects the reliability of devices. FEA

results in this study suggest the importance of the interfacial structure.

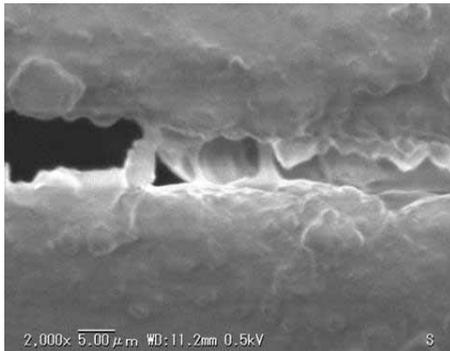

Fig.1 Interfacial structure between resins in the front of crack tip with imperfect bonding.

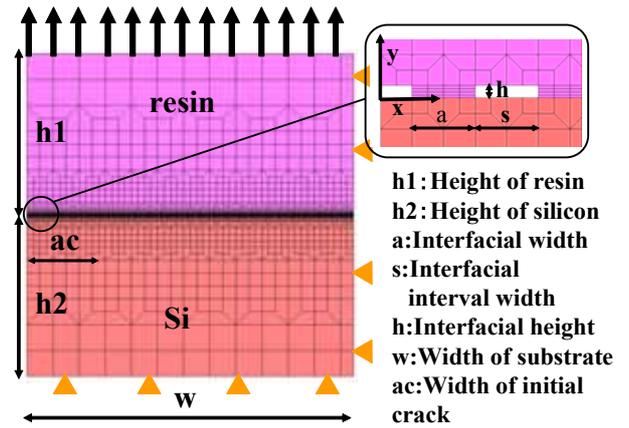

Fig.2. Proposed model

h1: Height of resin
h2: Height of silicon
a: Interfacial width
s: Interfacial interval width
h: Interfacial height
w: Width of substrate
ac: Width of initial crack

Table I Each parameter of model [μm]

| h1 | h2 | w | a | s | h | numbers of elements | numbers of nodes |
|---|---|---|---|---|---|---|---|
| 500 | 500 | 1000 | 5 | 5 | 1 | 7626 | 8372 |
| 500 | 500 | 1000 | 5 | 3 | 1 | 9532 | 10459 |
| 500 | 500 | 1000 | 5 | 1 | 1 | 10230 | 11154 |
| 500 | 500 | 1000 | 5 | 5 | 0.5 | 7626 | 8373 |
| 500 | 500 | 1000 | 5 | 5 | 2 | 7626 | 8372 |

Table II Material properties

| | Young's modulus (GPa) | Poisson's ratio (-) | CTE (*$10^{-6}$/K) | Initial yield stress (MPa) |
|---|---|---|---|---|
| resin | 10 | 0.3 | 19 | 60 |
| interface | 10 | 0.3 | 19 | 60 |
| silicon | 130 | 0.26 | 3.6 | - |

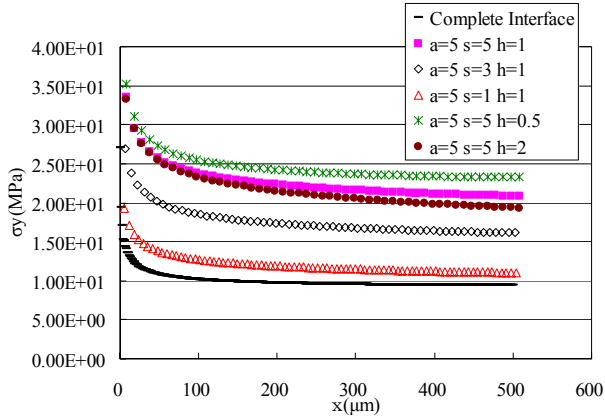

Fig.3.(a) Distributions of $\sigma_y$ when uniform tensile stress was applied on the upper of resin
(h1=h2=500 w=1000 μm)

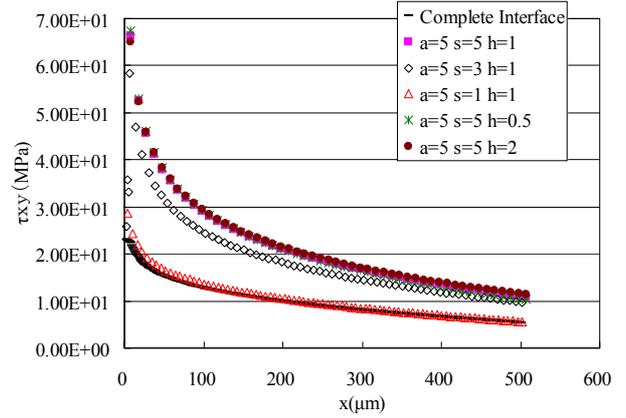

Fig.4. (b) Distribution of $\tau_{xy}$
(20C from 150C h1=h2=500μm w=1000μm)

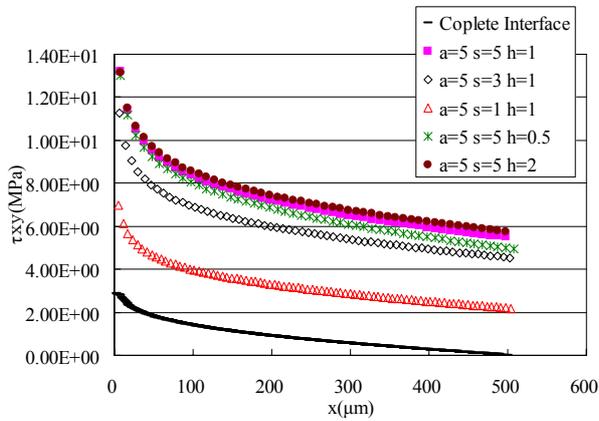

Fig.3. (b) Distribution of $\tau_{xy}$ when uniform tensile stress was applied on the upper of resin
(h1=h2=500μm    w=1000μm)

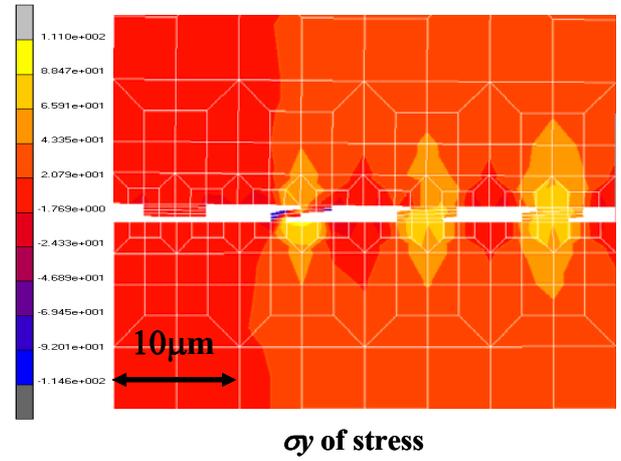

$\sigma y$ of stress

Fig.5. Crack growth from the crack tip

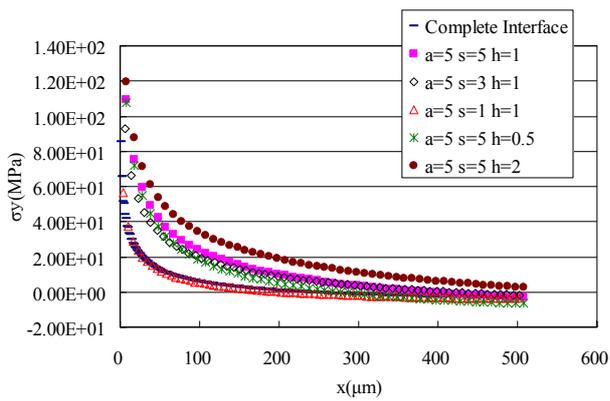

Fig.4. (a) Distribution of $\sigma_y$
(20C from 150C h1=h2=500 w=1000μm)

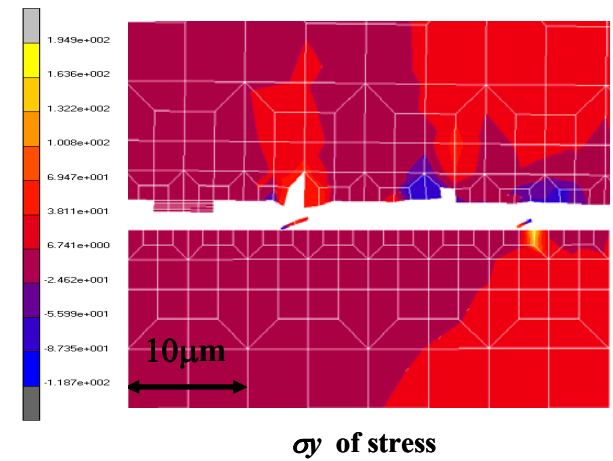

$\sigma y$ of stress

Fig.6. Crack growth from the crack tip

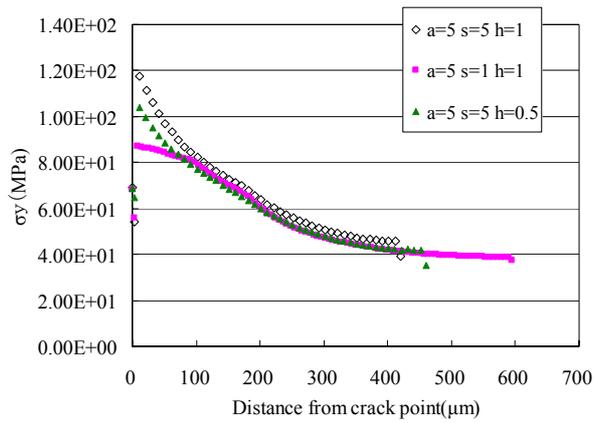

Fig.7. Distribution of $\sigma_y$ from the crack tip at the instability of fracture
(ac=100μm  h1=h2=500μm  w=1000μm )

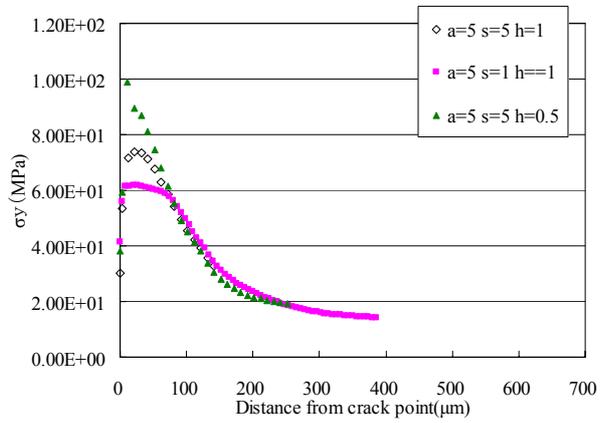

Fig.8. Distribution of $\sigma_y$ from the crack tip at the instability of fracture
(ac=300μm    h1=h2=500μm    w=1000μm )